\documentclass[a4paper,11pt]{JHEP3}
\usepackage{graphicx}
\usepackage{epsfig}
\usepackage{amsmath, amssymb}
\usepackage{dcolumn}% Align table columns on decimal point
\usepackage{bm}% bold math
\usepackage{amsfonts}

\usepackage{subfigure}
\usepackage{amssymb}

\def\dj{\hbox{d\kern-0.347em \vrule width 0.3em height 1.252ex depth
-1.21ex \kern 0.051em}}

\def\ran{\rangle}
\def\lan{\langle}

\newcommand{\ra}{\rightarrow}
\newcommand{\dd}{\partial}

\newcommand{\be}{\begin{equation}}
\newcommand{\ee}{\end{equation}}
\newcommand{\ben}{\begin{equation*}}
\newcommand{\een}{\end{equation*}}
\newcommand{\bea}{\begin{eqnarray}}
\newcommand{\eea}{\end{eqnarray}}
\newcommand{\bean}{\begin{eqnarray*}}
\newcommand{\eean}{\end{eqnarray*}}
\newcommand{\brr}{\begin{array}}
\newcommand{\err}{\end{array}}
\newcommand{\bc}{\begin{center}}
\newcommand{\ec}{\end{center}}
\newcommand{\nn}{\nonumber }

\newcommand{\gsim}{\,\raisebox{-0.6ex}{$\buildrel > \over \sim$}\,}

\newcommand{\bk}{{\mathbf k}}

\newcommand{\bp}{{\mathbf p}}

\newcommand{\bx}{{\mathbf x}}

\newcommand{\bq}{{\mathbf q}}

\newcommand{\CC}{{\cal C}}
\newcommand{\GG}{{\cal G}}
\newcommand{\HH}{{\cal H}}
\newcommand{\LL}{{\cal L}}

\newcommand{\PP}{{\cal P}}
\newcommand{\OO}{{\cal O}}
\newcommand{\RR}{{\cal R}}

\newcommand{\TT}{{\tt TT}}

\newcommand{\cd}{\cdot}

\newcommand{\de}{\delta}

\newcommand{\ep}{\epsilon}
\newcommand{\ga}{\gamma}
\newcommand{\Ga}{\Gamma}

\newcommand{\la}{\lambda}
\newcommand{\Om}{\Omega}

\newcommand{\lp}{\left}
\newcommand{\rp}{\right}

\title{Gravitational waves from self-ordering scalar fields}
\author{Elisa Fenu 
\\ Institute de Physique Th\'eorique, Universit\'e de Gen\`eve, 24 quai E. Ansermet, 1211 Gen\`eve 4, Switzerland\\ 
 \email{elisa.fenu@unige.ch} }
  \author{Daniel G. Figueroa
 \\ Instituto de F\'{\i}sica Te\'orica CSIC-UAM and Departamento de F\'isica Te\'orica,\\ 
 Universidad Aut\'onoma de Madrid, Cantoblanco 28049 Madrid, Spain
 \email{daniel.figueroa@uam.es}}
  \author{Ruth Durrer  
 \\ Institute de Physique Th\'eorique, Universit\'e de Gen\`eve, 24 quai E. Ansermet, 1211 
 Gen\`eve 4, Switzerland\\ 
  \email{ruth.durrer@unige.ch} }
 \author{Juan Garc\'{\i}a-Bellido
 \\ Instituto de F\'{\i}sica Te\'orica CSIC-UAM and Departamento de F\'isica Te\'orica,\\ 
 Universidad Aut\'onoma de Madrid,  Cantoblanco 28049 Madrid, Spain\\
PH-TH Department CERN, 1211 Gen\`eve 23, Switzerland\\
 \email{juan.garciabellido@uam.es} }

\received{\today}


\abstract{
Gravitational waves were copiously produced in the early Universe whenever the processes 
taking place were sufficiently violent. The spectra of several of these gravitational wave 
backgrounds on subhorizon scales have been extensively studied in the literature. In this paper 
we analyze the shape and amplitude of the gravitational wave spectrum on scales which are 
superhorizon at the time of production.  Such gravitational waves are expected from the self 
ordering of randomly oriented scalar fields which can be present during a thermal phase transition 
or during preheating after hybrid inflation. We find that, if the gravitational wave source acts only 
during a small fraction of the Hubble time, the gravitational wave spectrum at frequencies lower 
than  the expansion rate at the time of production behaves as $\Omega_{\rm GW}(f) \propto f^3$ 
with an amplitude much too small to be observable by gravitational wave observatories like LIGO, 
LISA or BBO. On the other hand, if the source is active for a much longer time, until a given mode 
which is initially superhorizon ($k\eta_* \ll 1$), enters the horizon, for $k\eta \gtrsim 1$, we find 
that the gravitational wave energy density is frequency independent, i.e. scale invariant. Moreover, 
its amplitude for a GUT scale scenario turns out to be within the range and sensitivity of BBO and 
marginally detectable by LIGO and LISA. This new gravitational wave background can compete 
with the one generated during inflation, and distinguishing both may require extra information.}

\preprint{IFT-UAM/CSIC-09-34, CERN-PH-TH/2009-145}

\maketitle

\begin{document}

%%%%%%%%%%%%%%%%%%%%%%%%%%%%%%%%%%%%%%%%%%%%%%
\section{Introduction}\label{introduction}
%%%%%%%%%%%%%%%%%%%%%%%%%%%%%%%%%%%%%%%%%%%%%%

Gravitational waves (GWs) are produced in the late Universe via cataclismic astrophysical 
events like hypernovae and inspiralling binaries. Because gravity is so weak, it is extremely 
difficult to detect directly with present day interferometers~\cite{GW1}. On the other hand, during 
the violent processes which  we expect took place in the very early Universe, several stochastic 
backgrounds of GWs of significant energy may be produced, although their amplitude today is 
drastically reduced by the expansion of the Universe, making them equally difficult to detect~\cite{GW2}. 
Their discovery  may however be possible in the near future, opening a completely new window into the uncharted territory of the very early Universe. For this we must determine the detailed GW spectrum, which strongly depends 
on the physical processes generating them.

In the last few years there has been significant  progress in the experimental prospects for detecting
GWs with interferometers like LIGO and VIRGO and the future satellite mission
LISA. This has stimulated research for sources of primordial GWs from
the early Universe, either from hypothetical first order phase 
transitions~\cite{ew1,Kam,ew2,ew3,Caprini:2007xq,Huber:2008hg} 
or from the  process of reheating after 
inflation~\cite{Khlebnikov97,GarciaBellido:1998,Easther1,Bellido1,Easther2,Bellido2,Dufaux1,Dufaux2}.

The mechanism responsible for GW production during these early Universe 
phenomena is typically a causal process, like bubble collisions or turbulence, giving rise
to spectra which peak at  wavelengths that are well within the causal horizon during their generation. 
Thus, most of past analyses concentrate on contributions 
of GWs with wavelengths smaller than the horizon at the time of production, 
with the exception of those 
generated during inflation~\cite{Starobinsky}, which are stretched by the inflationary expansion.

In this paper we study the infrared behaviour of the GW spectrum produced either  during preheating 
or during first order phase transitions, on scales which are superhorizon at 
the time of formation, {\it i.e.} $k < \HH_*$, where $k$ and $\HH_*$ are the comoving momentum and 
inverse horizon. We want to study a causal process of symmetry breaking 
like hybrid preheating~\cite{GarciaBellido:1997,GarciaBellido:1999,Felder:2000,GarciaBellido:2001,GarciaBellido:2002,DiazGil:2007}, where the order parameter has global $\mathcal{O}(N)$ symmetry in the false vacuum
and, upon symmetry breaking, the $N$ fields undergo self-ordering on a given scale as soon as they enter the horizon, in particular on scales much larger than the inverse mass of the field in the true vacuum.

We consider a multi-component scalar field which obtains a non-zero \textit{vacuum expectation 
value} ($vev$) $v$ and a mass $m$, during a symmetry breaking process. We shall assume that this 
mass $m$ is much larger than the Hubble parameter $H_*$ at the time of the transition, since if the 
{\em vev} in the true vacuum is much smaller than the Planck scale, then $H_* \sim m\,v/M_p \ll m$. 
Such a model could describe the symmetry breaking process which triggers the end of hybrid inflation 
or a thermal phase transition. 
As long as we are only interested in superhorizon scales, $k\gg \HH_*$, we 
can neglect the radial, massive mode and treat the dynamics within the non-linear 
sigma-model (NLSM) approximation. On large scales, the
anisotropic stresses are determined by gradient energy and the typical 
(comoving) scale is simply the time dependent horizon scale $\HH^{-1}$. The 
field self-orders at the horizon scale, and
the source of GWs decays inside the horizon. For scalar metric 
perturbations this process has been studied {\it e.g.} in Ref.~\cite{DK}. 
It is very closely related to the scaling of global topological 
defects~\cite{DKM} even though for a number of components $N>4$ there are no 
topological defects associated with such a scalar field in $3+1$ dimensions. 

We work in the large $N$ approximation within which the scalar field equation 
of motion, for scales larger than the inverse mass, $k \ll m$, 
can be solved analytically. The GW spectrum will then be estimated by 
analytical approximations, introducing the anisotropic stress tensor sourced by 
the field fluctuations at different scales. 

Tensor perturbations from a NLSM in the large $N$ approximation have also been 
studied in Ref.~\cite{Krauss:1991,JonesSmith:2007ne}, see also~\cite{new}. There the authors 
have calculated the tensor perturbation spectrum for scales which enter the horizon in the 
matter era and they have compared this with the inflationary signal in the CMB.
Here we shall concentrate on the radiation dominated era and the detection of 
the signal in direct gravitational wave experiments like advanced 
LIGO~\cite{ligo}, LISA~\cite{lisa} and BBO~\cite{bbo}.

The paper is organized as follows. In the next section we describe the 
formalism, derive the scalar field solutions and calculate the unequal time 
anisotropic stress correlators which source GWs.
In Section~\ref{s:GWprod} we study the production of GWs from long wavelength modes 
of this source. We derive a general formula that 
can be applied to different situations, depending how long the GW source is acting. 
In Section~\ref{s:GWtoday} we use this result to
determine the shape and amplitude of the GW spectrum 
in two situations, first the case of a source producing GWs only during a small 
fraction of the Hubble time and, second, the case in which the source producing
GWs acts for a much longer time, until  a given mode which is initially superhorizon, 
$k\eta_* \ll 1$, enters the Hubble radius,
$k\eta \simeq 1$. In Section~\ref{s:con} we summarize our results and conclude.
\\

{\bf Notation} Throughout this paper we assume a spatially flat Friedmann
Universe with metric
\be
ds^2 =a^2(\eta)\left(-d\eta^2 +\de_{ij}dx^idx^j\right)\, ,
\ee
where $\eta$ denotes conformal time and we normalize the scale factor to 
unity today, $a(\eta_0)=1$. The comoving Hubble rate is $\HH =a'/a$, 
while $H=a'/a^2$ is the physical one. The prime denotes derivative w.r.t. 
conformal time $\eta$.

\section{Formalism}\label{s:form}

We first introduce the NLSM and the large $N$ limit of a 
global $\OO(N)$ symmetric scalar field, then we study the physics of the correlators 
of the anisotropic stress tensor. 

\subsection{The model}

We consider an $N$-component scalar field with a Lagrangian
\be
\LL = \LL_0 + \LL_1=  -\dd_\mu\Phi^{\rm T}\dd^\mu\Phi 
-\la\left(\Phi^{\rm T}\Phi-\frac{v^2}{2}\right)^2  + \LL_1\ ,
\ee
where $\Phi^{\rm T} = (\phi_1,\phi_2,...,\phi_N)/\sqrt{2}$, $\lambda$ is the dimensionless 
self-coupling of $\Phi$ and $v$ is the $vev$ in the true vacuum. In the case of a thermal 
bath at high temperature, the Lagrangian $\LL_0$ obtains corrections of the form 
$\LL_1 \sim -T^2\Phi^2$, so that its minimum is at $\Phi=0$ which respects the global $\OO(N)$ 
symmetry of the Lagrangian. At low temperature, $T < T_c \simeq v$, the
thermal corrections are too small to the keep the minimum at $\Phi=0$ and the 
global $\OO(N)$ symmetry is spontaneously broken to $\OO(N-1)$. 
In the context of hybrid preheating, there is no need for thermal restoration of the symmetry. 
The field $\Phi$ acquires a large mass during inflation through its coupling to the inflaton 
$\chi$, $\LL_1 = -g^2\Phi^{\rm T}\Phi\chi^2$. Above a critical value, 
$\chi > \chi_c \equiv \sqrt{\lambda}v/g$, the effective quadratic mass of $\Phi$ is positive and the field 
is fixed at $\Phi = 0$. 
When the quadratic mass becomes negative, $\chi < \chi_c$, a tachyonic instability triggers the end of inflation and symmetry breaking. Soon after the symmetry is broken, 
thermal corrections and tachyonic effects can be neglected, and $\Phi$ is closely confined 
(in most of space) to the vacuum manifold, given by $\sum_a\phi_a^2({\bf x},\eta) = v^2$. 
Nevertheless, in positions such that their comoving distance is 
$|\bx-\bx'|>\HH^{-1}$, the values $\Phi(\bx,\eta)$ and $\Phi(\bx',\eta)$ are
uncorrelated, which leads to a gradient energy density associated to 
the $N-1$ Goldstone modes, $\rho \sim (\partial_i\Phi)^2$. For $N > 2$, the 
dynamics of the Goldstone modes is well described by a NLSM~\cite{TS,DKM} where we force $\sum_a\phi_a^2 = v^2$ by a Lagrange 
multiplier. This corresponds to the limit $\la\ra\infty$ in the above 
Lagrangian. This approximation is very good for physical scales with are much 
larger than $m^{-1} \equiv 1/(\sqrt{\la}v)$. Of course, on small scales the field 
fluctuations still oscillates around the true \textit{vev}, but in this paper we only 
focus on the superhorizon modes which are free to wander around in the vacuum manifold, 
giving rise to a gradient energy density which will generate GWs on these scales.

Normalizing the symmetry breaking field to its $vev$, $\beta \equiv \Phi/v$, each component of 
the field obeys the non-linear sigma model evolution equation~\cite{DK}
 \begin{eqnarray}\label{e:sigma}
\Box\beta^a -(\dd_\mu\beta\cd\dd^\mu\beta)\beta^a &=&0  ~,
\end{eqnarray}
where $(\dd_\mu\beta\cd\dd^\mu\beta) =\sum_a\eta^{\mu\nu}\dd_\mu\beta^a(\mathbf{x},\eta)
\dd_\nu\beta^a(\mathbf{x},\eta)$ and $\sum_a\beta^a(\mathbf{x},\eta)
\beta^a(\mathbf{x},\eta) = 1$. In the large $N$-limit, we assume that the sum 
over components can be replaced by an ensemble average,
\begin{eqnarray}
T(x) = \sum_a \eta^{\mu\nu}\partial_\mu\beta^a\partial_\nu\beta^a = 
N \langle \eta^{\mu\nu}\partial_\mu\beta^a\partial_\nu\beta^a \rangle = \bar T(\eta) \ .
\end{eqnarray}
By dimensional considerations, $T \propto \mathcal{H}^2$, or
\begin{eqnarray}
\bar T(\eta) = T_o\eta^{-2} \,,
\end{eqnarray}
with $T_o>0$. Replacing the non-linearity in the sigma-model by this expectation value
we obtain a linear equation which can be solved exactly. In Fourier space 
it reads
\begin{eqnarray}\label{e:sigma2}
\beta_k^{a\,''} + \frac{2\gamma}{\eta}\beta_k^{a\,'}
  + \left(k^2-\frac{T_o}{\eta^2}\right)\beta_k^a = 0\,,
\end{eqnarray}
where $\gamma = d\log a/d\log\eta$ and primes denote derivatives 
w.r.t. $\eta$. In a radiation dominated 
Universe $\ga=1$ while in a matter dominated Universe $\ga=2$. 
The solution to Eq.~(\ref{e:sigma2}) for constant $\ga$ is given by
\begin{eqnarray}
\beta^a(\bk,\eta) = (k\eta)^{\frac{1}{2}-\gamma}\Big[C_1(\bk)\,J_\nu(k\eta)+C_2(\bk)\,Y_\nu(k\eta)\Big]\,,
\end{eqnarray}
where
\begin{eqnarray}\label{nu}
\nu^2 = \left(\frac{1}{2}-\gamma\right)^2+T_o \ ,
\end{eqnarray}
and $C_1$, $C_2$ are constants of integration.
Thus, $\nu > 1/2$ for a radiation dominated Universe and $\nu > 3/2$ for matter domination. 
Since in general we have that $\nu > 0$, $Y_\nu$ diverges for small argument, so we will keep only the regular mode of the solution $J_\nu$, which can be written as
\begin{eqnarray}\label{e:beta}
\beta^a(\bk,\eta) &=& \sqrt{A}\left(\frac{\eta}{\eta_*}\right)^{{1\over2}-\gamma}
\frac{J_\nu(k\eta)}{(k\eta_*)^\nu} \beta^a(\bk,\eta_*)\,,
\end{eqnarray}
where $\beta^a(k,\eta_*)$ is the $a$-th component of the field at the initial 
time $\eta_*$. We assume that $\beta$ is initially Gaussian distributed with 
a scale-invariant spectrum on large scales and vanishing power on small scales
\be\label{e:coreta*}
 \langle\beta^a(\bk,\eta_*)\beta^{*b}(\bk',\eta_*)\rangle = \left\{ 
\begin{array}{cl} 
 (2\pi)^3\CC\frac{\de^{ab}}{N}\de(\bk-\bk') & ,\,\, k\eta_* \ll 1 \\
  0  & ,\,\,k\eta_* >1 \ . \end{array} \right.
\ee
This means that the field is aligned on scales smaller than the comoving
horizon $\eta_*$ and has arbitrary orientation on scales larger than $\eta_*$. The
condition that $\beta^2=1$ actually introduces correlations between the different
components of $\beta$ but these lead to corrections of order $1/N$ to the 
above expression which we will neglect here. We also do not enter into the 
details of the decay of this function around $k\eta_*=1$. The constant 
$\CC$ is chosen such that the normalization condition is satisfied (up to
corrections of order $1/N$),
\bea 
\beta^2(\bx,\eta_*) &\equiv&
\langle\beta^2(\bx,\eta_*)\rangle \Big(1 +\OO(1/N)\Big) 
 \nn \\&\simeq& \int\frac{d^3k}{(2\pi)^3} 
\frac{d^3k'}{(2\pi)^3}  \langle\beta^a(\bk,\eta_*)\beta^{*a}(\bk',\eta_*)\rangle e^{i\bx\cdot(\bk-\bk')}   
\simeq   \frac{\CC}{6\pi^2\eta_*^3}  =1 \,.
\eea
In the large $N$-limit we neglect the corrections of order $1/N$ which 
come from the fluctuations in $\beta^2$. On large scales this is a very good 
approximation. However, on small scales, and in particular, on scales 
comparable with the inverse of the mass of the symmetry breaking field, $m^{-1}$,
the fluctuations are certainly not negligible.  In our 
analysis we consider only large scales, where the above approximation is valid.

In order for $\langle\beta^2\rangle$ to be time independent we need that
the equal time correlator be fixed to one:
\bea
\langle\beta^2(\bk,\eta)\rangle &=& A\,\CC\int\frac{d^3k}{(2\pi)^3}
\left(\frac{\eta}{\eta_*}\right)^{(1-2\gamma)}
\frac{J^2_\nu(k\eta)}{(k\eta_*)^{2\nu}}
\nn \\ \label{eq:beta2}  
 &\simeq&3A\left(\frac{\eta_*}{\eta}\right)^{2(1+\gamma-\nu)}
 \int_0^{\infty} dy y^{2(1-\nu)}J_\nu^2(y) \,=\, 1\,, 
 \eea
where we have substituted $\CC =6\pi^2\eta_*^3$ and we have set 
$y=k\eta$. Note that the upper limit is actually $\eta/\eta_*$, but at late times, 
the (dimensionless) integral is insensitive to the upper boundary, so we can take 
it to infinity and thus make the integral free of any time scale. In order 
to obtain a time-independent $vev$, we then just require
\be
\nu =\gamma+1 \ .
\ee
Introducing this relation into Eq.~(\ref{nu}), one obtains $T_o$ in terms of $\gamma$ as
\bea
T_o &=& 3(\ga +1/4) \ .
\eea
The constant $A$ is determined then by the condition
\be\label{e:A}
  1 =  3A\int_0^\infty dy y^{2(1-\nu)}J_\nu^2(y)\ , \qquad  {\rm hence} \qquad 
  A = \frac{4\Ga(2\nu-1/2)\Ga(\nu-1/2)}{3\Ga(\nu-1)} \ .
\ee
Since $\nu = \ga+1$, we can also write the amplitude of the field fluctuations, as
\be \label{e:beta2}
\beta^a(\bk,\eta) =\sqrt{A}\left(\frac{\eta}{\eta_*}\right)^{3/2}
   \frac{J_\nu(k\eta)}{(k\eta)^\nu}\beta^a(\bk,\eta_*) \ .
\ee

\subsection{Unequal time correlators}

{}From Eqs.~(\ref{e:coreta*})~and~(\ref{e:beta2}) we obtain the following 
expression for the unequal time correlator of the field: 
\begin{eqnarray}
\left\langle \beta^a(\mathbf{k},\eta)\beta^{*b}(\mathbf{k}',\eta')\right\rangle &=& 
A\left(\frac{\eta\eta'}{\eta_*^2}\right)^{3/2}
  \frac{J_\nu(k\eta)J_\nu(k'\eta')}{(k\eta)^\nu(k'\eta')^\nu}\,
 \left\langle\beta^a(\mathbf{k},\eta_*)\beta^{*b}(\mathbf{k}',\eta_*)\right\rangle \nn\\
&=& (2\pi)^3 6\pi^2A(\eta\eta')^{3/2}\,
\frac{J_\nu(k\eta)J_\nu(k\eta')}{(k\eta)^\nu(k\eta')^\nu}\,
\frac{\de_{ab}}{N}\de(\bk-\bk') \nn\\
&\equiv& (2\pi)^3\de(\bk-\bk')\PP^{ab}_\beta(k,\eta,\eta')\,.
\end{eqnarray}

We assume that the field $\beta$ is Gaussian distributed initially. As its
time evolution is linear, it will remain a Gaussian field and we can 
determine higher order correlators via Wick's theorem. This will be important 
in the next section when we determine the unequal time correlator of the 
anisotropic stresses which source the production of GWs. 
Furthermore, this source is {\em totally coherent}~\cite{DKM} in the sense
that its unequal time correlator $\PP^{ab}_\beta(k,\eta,\eta')$ is a product 
of a function of $\eta$ and $\eta'$,
\bea
\label{e:Pbeta}
\PP^{ab}_\beta(k,\eta,\eta') &=& \frac{\de_{ab}}{N} 6\pi^2A(\eta\eta')^{3/2}
\frac{J_\nu(k\eta)J_\nu(k\eta')}{(k\eta)^\nu(k\eta')^\nu} \ \equiv\ \frac{\de_{ab}}{N}
f(k,\eta)f(k,\eta') ~,  \label{e:unequal}\\
\mbox{with} ~~f(k,\eta) &=& \sqrt{6\pi^2A}\ k^{3/2}\frac{J_\nu(k\eta)}{(k\eta)^{\nu-3/2}} \nn
\ .
\eea
Note the $k^{3/2}$ scaling law at horizon crossing ($k\eta\sim1$) which is characteristic for
quantum fluctuations from de Sitter, {\it i.e.} inflation. This already hints to the fact that we will 
find a scale-invariant spectrum also in this case. 

\section{The production of gravitational waves}\label{s:GWprod}

In this section we  derive a general formula for the GW power spectrum sourced by superhorizon modes of a self ordering field. 
We also comment about the frequency range for the GW background
produced in this way. 

%\subsection{Gravitational wave spectrum}

Let us consider tensor perturbations (GWs) of the metric, 
\begin{eqnarray}
ds^2 = a^2(\eta)(\eta_{\mu\nu} + 2h_{\mu\nu})dx^{\mu}dx^{\nu} \ ,
\end{eqnarray}
where $h_{ij}$ is traceless, $h^i_i=0$, and divergence free, $\partial^i h_{ij}=0$.
Linearizing Einstein's equations yields the evolution equation of GWs sourced by 
the anisotropic stresses of the scalar fields $\Phi$,
\be\label{e:h''}
h_{ij}''(\mathbf{x},\eta) + 2\mathcal{H} \, h_{ij}'(\mathbf{x},\eta) - 
\nabla^2h_{ij}(\mathbf{x},\eta) = 8\pi G\,\Pi_{ij}(\mathbf{x},\eta) \ ,
\ee
where $\Pi_{ij}$ represents the \TT\ part of 
the (effective) anisotropic stress tensor
\begin{eqnarray}
T_{ij}(\mathbf{x},\eta) = \partial_i\phi^a(\mathbf{x},\eta)
  \partial_j\phi^a(\mathbf{x},\eta) -\frac{1}{3}\de_{ij}
  \lp[\nabla\phi^a(\mathbf{x},\eta)\rp]^2 \,.
\end{eqnarray}

Fourier transforming the GW evolution equation (\ref{e:h''}) we obtain
\begin{eqnarray}
h_{ij}''(\mathbf{k},\eta) + 2\mathcal{H} \, h_{ij}'(\mathbf{k},\eta) + k^2h_{ij}(\mathbf{k},\eta) = 8\pi G\,\Lambda_{ij,lm}(\hat \bk)T_{lm}(\mathbf{k},\eta)
\end{eqnarray}
where the projector 
\begin{eqnarray*}
&&\Lambda_{ij,lm}(\hat \bk) \equiv P_{il}(\hat \bk)P_{jm}(\hat \bk) - \frac{1}{2}P_{ij}(\hat \bk)P_{lm}(\hat \bk)~,\\
&& P_{ij}(\hat \bk) \equiv \delta_{ij} - \hat \bk_i\hat \bk_j~, \qquad \hat\bk \equiv \bk/k  ~,
\end{eqnarray*}
filters out the \TT\ part of the Fourier transformed effective anisotropic 
stress tensor
\begin{eqnarray}
\Pi_{ij}(\mathbf{k},\eta) = \Lambda_{ij,lm}(\hat \bk)\int \frac{d^3q}{(2\pi)^3} \,
q_lq_m\, \phi^a(\mathbf{q},\eta)  \phi^a(\mathbf{k-q},\eta)\ .
\end{eqnarray}
Note that we are summing over repeated indices both in coordinates 
and in field components.

The 2-point correlation function of the tensorial part of the anisotropic
 stress-tensor is of the form
\begin{eqnarray}
\Big\langle \Pi_{ij}(\bk,\eta)\Pi_{lm}^*(\bk',\eta')\Big\rangle \equiv 
 (2\pi)^3\delta(\bk-\bk')\Pi^2(k,\eta,\eta')\mathcal{M}_{ijlm}(\hat\bk)~,
\end{eqnarray}
where
\begin{eqnarray}
\mathcal{M}_{ijlm}(\hat \bk) = \frac{1}{4}\lp[\Lambda_{ij,lm}(\hat \bk) + \Lambda_{ij,ml}(\hat \bk)\rp] \ .
\end{eqnarray}
Since the trace $\mathcal{M}_{ijij}=1$,
\be
\label{e:Pi}
\Big\langle \Pi_{ij}(\bk,\eta)\Pi_{ij}^*(\bk',\eta')\Big\rangle \equiv 
 (2\pi)^3\delta(\bk-\bk')\Pi^2(k,\eta,\eta')~.
\ee
To determine $\Pi^2(k,\eta,\eta')$, we compute 
$\Big\langle \Pi_{ij}(\bk,\eta)\Pi_{ij}^*(\bk',\eta')\Big\rangle$ explicitly using 
Wick's theorem to reduce 4-point functions 
of the field to products of 2-point functions
\begin{eqnarray}
& & \Big\langle \Pi_{ij}({\mathbf
    k},\eta)\Pi_{lm}^*(\mathbf{k'},\eta')\Big\rangle = \nn \\
&=&
    \Lambda_{ij,pq}(\hat \bk)\,\Lambda_{lm,rs}(\hat \bk')\int\,\frac{d^3q}{(2\pi)^3}
    \frac{d^3q'}{(2\pi)^3}\,q_pq_qq_r'q_s'\Big\langle\phi^a(\mathbf{q},\eta)
 \phi^a(\mathbf{k-q},\eta)\phi^{*b}(\mathbf{q}',\eta')\phi^{*b}(\mathbf{k-q},\eta')\Big\rangle \nn \\
&=& \int \frac{d^3q\,d^3q'}{(2\pi)^6}\left( q^{\rm T}
\Lambda q\right)_{ij}\left({q'}^{\rm T}\Lambda q'\right)_{lm}
\left[\Big\langle\phi^a(\mathbf{q},\eta)\phi^{*a}(\mathbf{q-k},\eta)
\rangle\langle\phi^b(\mathbf{-q}',\eta')\phi^{*b}(\mathbf{k'-q'},\eta')\Big\rangle
  \right. + \nn \\ && \! \left. \quad
+\ \Big\langle\phi^a(\mathbf{q},\eta)\phi^{*b}(\mathbf{q}',\eta')\Big\rangle
\Big\langle\phi^a(\mathbf{k-q},\eta)\phi^{*b}(\mathbf{k'-q'},\eta')\Big\rangle 
\right. + \nn \\ && \! \left. \quad
+ \Big\langle\phi^a(\mathbf{q},\eta)\phi^{*b}(\mathbf{k'-q'},\eta')\Big\rangle
\Big\langle\phi^a(\mathbf{k-q},\eta)\phi^{*b}(\mathbf{q}',\eta')\Big\rangle\right] \nn
\\ &=& \int d^3q\,d^3q'\left( q^{\rm T}\Lambda q\right)_{ij}
\left( {q'}^{\rm T}\Lambda q'\right)_{lm}   \Big[\PP_\phi^{aa}(|\mathbf{q}|,\eta,\eta)
  \PP_\phi^{bb}(|\mathbf{q}'|,\eta',\eta')\,\delta(\bk)\delta(\bk')    \nn  \\
&& \quad+ \
  \PP_\phi^{ab}(|\mathbf{q}|,\eta,\eta')\PP_\phi^{ab}(|\mathbf{k-q}|,\eta,\eta')\,
 \delta(\mathbf{q-q'})\delta(\mathbf{k-q}-\mathbf{k'+q'})  \nn \\ 
 \label{e:4point}
&& \quad  + \
  \PP_\phi^{ab}(|\mathbf{q}|,\eta,\eta')\PP_\phi^{ab}(|\mathbf{k-q}|,\eta,\eta')\,
 \delta(\mathbf{q'+q-k'})\delta(\mathbf{q'+q-k})\Big] \,
\end{eqnarray}
where we use the notation $\left( q^{\rm T}\Lambda q\right)_{ij} \equiv 
 q_l\Lambda_{ij,lm}q_m$ and we have introduced the reality condition $\phi^*(\bk)=\phi(-\bk)$ 
 and the unequal time correlator 
of the field $\phi$ which is defined in the same way as the one for $\beta$,
\be
\langle\phi^a(\bk,\eta)\phi^{*b}(\bk',\eta')\rangle =(2\pi)^3\de(\bk-\bk')\,\PP_\phi^{ab}(k,\eta,\eta')\ .
\ee
The zero-mode of the anisotropic stresses vanishes 
due to isotropy so that the first term in the square bracket of the
integral (\ref{e:4point}) does not contribute.

We now can compute the unequal time correlator
$\left\langle \Pi_{ij}(\bk,\eta)\Pi_{ij}^*(\bk',\eta')\right\rangle$. Using
\begin{eqnarray}
\left( q^{\rm T}\Lambda q\right)_{ij}\left(q^{\rm T}\Lambda q\right)_{ij} = 
\frac{1}{2}q^4\left(1-(\hat\bk\cdot\hat\bq)^2\right)^2\ , 
\end{eqnarray} 
we obtain
\begin{eqnarray}\label{e:Pi2}
\Pi^2( k,\eta,\eta')= 
\int \frac{d^3q}{(2\pi)^3} \,  q^4
 \left[1-(\hat\bk\cdot\hat\bq)^2\right]^2\PP_\phi^{ab}(|\mathbf{q}|,\eta,\eta')
\PP_\phi^{ab}(|\mathbf{k-q}|,\eta,\eta') \ .
\end{eqnarray} 

We now relate the GW energy density spectrum to the unequal time anisotropic stress 
spectrum of the source, $\Pi^2( k,\eta,\eta')$. For this we first write the GW evolution equation in momentum space, 
\be
   h_{ij}'' + 2\frac{ a'}{a}  h'_{ij} +  k^2h_{ij} = 8 \pi G \,\Pi_{ij} ~.
\ee
Defining a new variable $\bar{h}_{ij} \equiv a h_{ij}$, one obtains
\be
  {\bar{h}}''_{ij} +\lp(k^2-\frac{ a''}{a} \rp) \bar{h}_{ij} = 8 \pi G  a \, \Pi_{ij}  ~.
\ee
In a radiation dominated background ($a \propto \eta$) this reduces to
\be
  \label{eq:hbar}
  {\bar{h}}''_{ij} +k^2 \bar{h}_{ij} = 8 \pi G   a \, \Pi_{ij}  ~.
\ee
The solution of this differential equation with the initial conditions 
$h_{ij} = h_{ij}' = 0$ is given by the convolution of the source with the 
 Green function $\GG(k,\eta,\eta') =\sin (k\eta-k\eta')$, 
\be
  \label{e:nonfreeGW}
  \bar{h}_{ij}(\bk,\eta<\eta_{\rm fin})= \frac{8\pi G}{k^2}\int_{x_*}^{x} dy ~a(y/k) \,\Pi_{ij}(\bk,y/k) \sin (x-y) \,,
\ee 
where we have set $x \equiv k\eta$ and $y \equiv k\eta'$.
The source of gravity waves is acting for a time interval 
$\delta\eta_* = (\eta_{\rm fin}-\eta_*) 
= \epsilon\eta_*$. If $\epsilon < 1$ we call the process short-lasting.
This is the relevant case for example for GWs produced during 
a symmetry breaking phase transition where the source disappears after the 
phase transition since the latter typically lasts only for a fraction of the Hubble time. However, the Goldstone modes considered in this work may 
very well be long lived as they are not expected to interact with ordinary 
matter. In this case therefore a long lasting source may be better motivated.
We discuss both cases below.

After the source has decayed, GWs are freely propagating, 
and thus described by the homogeneous solution of Eq.~(\ref{eq:hbar}),
\be
 \label{e:freeGW}
  \bar{h}_{ij}(\bk,\eta>\eta_{\rm fin})=
            A_{ij}(\bk)\sin (k\eta-k\eta_{\rm fin})+
            B_{ij}(\bk)\cos (k\eta-k\eta_{\rm fin}) \,.
\ee
The coefficients $A_{ij}$ and $B_{ij}$ are fixed by matching the homogeneous 
solution to the inhomogeneous one at  $\eta=\eta_{\rm fin}$. Matching both $\bar{h}_{ij}$ and its derivative ${\bar{h}}'_{ij}$ yields 
\bea
&&  A_{ij}(\bk) = \frac{8\pi G}{k^2}\int_{x_*}^{x_{\rm fin}} dy ~a(y/k) \Pi_{ij}(\bk,y/k) \cos (x_{\rm fin}-y)  ~, 
     \nn \\ 
&&  B_{ij}(\bk) = \frac{8\pi G}{k^2}\int_{x_*}^{x_{\rm fin}} dy ~a(y/k) \Pi_{ij}(\bk,y/k) \sin(x_{\rm fin}-y)  ~.
\eea

The GW energy density is given by (see {\it e.g.}~\cite{ew3})
\be\label{e:rhoGW}
  \frac{d\rho_{\rm GW}}{d\log k}= \frac{k^3 |h'|^2(k,\eta)}
     {2(2\pi)^3Ga^2} ~,
\ee
where the GW power spectrum has been normalized as follows:
\be
  \Big\lan h'_{ij}(\bk,\eta) h'^*_{ij}(\bq,\eta)\Big\ran = 
  2\Big\lan h'_{+}(\bk,\eta) h'^*_{+}(\bq,\eta) +  h'_{\times}(\bk,\eta) 
   h'^*_{\times}(\bq,\eta)\Big\ran = 
(2\pi)^3 \delta^3(\bk-\bq) \,
     |h'|^2(k,\eta)  ~.
\ee
Here our normalization differs from that of Ref.~\cite{JonesSmith:2007ne}.
Their definition of the power spectrum is related to ours by
\be \label{def:P}
 {\cal P}(k,\eta) \equiv 2\pi k^3 |h|^2(k,\eta)
\ee 
and they infer
$ \frac{d\Om_{GW}(k,\eta_0)}{d\log k} = \frac{k^2{\cal P}(k,\eta)}{6H_0^2}$ 
whereas we obtain, with (\ref{e:rhoGW}) and $h'=kh$ for sub-horizon modes,
$$ \frac{d\Om_{GW}(k,\eta_0)}{d\log k} = \frac{k^5 |h|^2(k,\eta)}
     {6\pi^2H_0^2} = \frac{k^2{\cal P}(k,\eta)}{12\pi^3H_0^2}~. $$
This difference in the normalization, which we attribute to an error in 
Ref.~\cite{JonesSmith:2007ne}, leads to a reduction of the final result
by about a factor 60, which may be relevant for observations.

With the solution for $\bar h_{ij}$ above, we obtain for $\eta>\eta_{\rm fin}$
\bea
  && |h'|^2(k,\eta) = \frac{1}{2a^2}\Big(k^2+\HH^2\Big)\Big(\lan A_{ij}A_{ij}^*\ran +
      \lan B_{ij}B_{ij}^*\ran \Big) 
      \nn \\ 
      && \quad= \frac{k^2+\HH^2}{2a^2} \lp(\frac{8\pi G}{k^2}\rp)^2 
      \int_{x_*}^{x_{\rm fin}}  dy \int_{x_*}^{x_{\rm fin}}  dz  ~a\lp(\frac{y}{k}\rp) 
      a\lp(\frac{z}{k}\rp)\cos(z-y)\Pi^2\lp(k,\frac{y}{k},\frac{z}{k}\rp),
\eea
where we have used Eq.~(\ref{e:Pi}).
The GW energy density at time $\eta$ is of course well 
defined only for waves with a wavelength well within the horizon, $k\gg\HH$.
Therefore we shall approximate $k^2+\HH^2 \simeq k^2$ in the following.

The GWs are sourced by the anisotropic stress of the scalar
field $\phi^a=v \beta^a$. The correlators are simply  related by
$$   \PP^{ab}_{\phi}= v^2 \PP^{ab}_{\beta}  \,.$$
With Eq.~(\ref{e:Pi2}) we obtain the following expression for the GW energy density after the decay of the source, $\eta>\eta_{\rm fin}$,
\begin{eqnarray}
  \frac{d\rho_{\rm GW}(k,\eta)}{ d\log k} 
 	&=& \frac{G\,v^4}{4\pi^4}\frac{k^3}{a^4(\eta)}
 	\int_{\eta_*}^{\eta_{\rm fin}}  d\tau   \int_{\eta_*}^{\eta_{\rm fin}} 
       d\xi \ a(\tau) a(\xi) \, \cos(k\xi-k\tau) \nn \\
    	&& \times\int d^3p~p^4 \sin^4\theta \
     	\PP_{\beta}^{ab}(p,\tau,\xi)\, \PP_{\beta}^{ab}(|\bk-\bp|,\tau,\xi)\,,
\end{eqnarray} 
where $\cos\theta \equiv \hat\bk\cdot\hat\bp$.
Inserting the power spectrum of $\beta$ in the above expression and 
summing over the field components, we find
\bea
\label{e:masterEq}
&&  \frac{d\rho_{\rm GW}(k,\eta)}{ d\log k}= \frac{G\,v^4}{4\pi^4}
  		\frac{k^3}{a^4(\eta)} \frac{36\pi^4A^2 }{N}
		\int_{\eta_*}^{\eta_{\rm fin}} d\tau    \int_{\eta_*}^{\eta_{\rm fin}}  d\xi \
     		a(\tau) a(\xi) \, \cos(k\xi-k\tau)  \nn\\ && \quad    \label{spectrumGW}
     		\times  \int_{\begin{tiny}\begin{array}{c}p<1/\eta_*\\|\bk-\bp|<1/\eta_*\end{array}\end{tiny}} d^3 p~p^4 \sin^4\theta ~
     		\tau^3 \xi^3 \
     		\frac{J_{\nu}(p\tau)}{(p\tau)^{\nu}} 
     		\frac{J_{\nu}(p\xi)}{(p\xi)^{\nu}}
     		\frac{J_{\nu}(|\bk-\bp|\tau)}{(|\bk-\bp|\tau)^{\nu}}
     		\frac{J_{\nu}(|\bk-\bp|\xi)}{(|\bk-\bp|\xi)^{\nu}} \,.
\eea

Here the constant $A$ comes from the normalization of $\beta$, and it is given by
Eq.~(\ref{e:A}). In the radiation dominated background considered here, we 
have $\nu=1+\ga=2$ and $A=5\pi/4$. Note also that we choose the normalization of the
scale factor such that $a(\eta_0)=1$. Hence the {\em comoving} wave number $k$ is simply
related to the present frequency of the GW by 
$$f= \frac{k}{2\pi}\ . $$ 

In the next section we evaluate the present amplitude and frequency 
dependence of the GW spectrum generated in this way explicitly. 
For this, the following relation between temperature and time in a radiation
dominated Universe are useful~\cite{mybook},
\be
 H^2(t) = \frac{1}{\eta^2\,a(\eta)^2} = 
\frac{8\pi G}{3}{\pi^2\over30}\,g_{\rm eff}(\eta)T^4(\eta)\,.
\ee
Assuming an adiabatic expansion, $g_{\rm eff}(aT)^3 =$ const., one finds
\be
 \eta = \frac{M_{\rm Pl}}{T(\eta)T_0}\left(\frac{g_{\rm eff}(\eta)}{2}
        \right)^{1/3}\left(\frac{45}{4\pi^3g_{\rm eff}(\eta)}\right)^{1/2}
  =1.6\times 10^7{\rm sec}\left(\frac{\rm GeV}{T}\right)g^{-1/6}_{\rm eff}(T) \ .
\ee
On the other hand, the expression for the temperature associated to a global 
$\OO(N)$ symmetry breaking is~\cite{ruth}
\be
T_* = \sqrt{\frac{24}{N+2}}v \ ,
\ee 
independent of the coupling $\la$. 

Before moving to the evaluation of Eq.~(\ref{spectrumGW}), let us briefly determine the
frequencies for the GW sources discussed in this paper. 
We are studying the IR modes $k\eta_* < 1$ 
of the GW spectrum, corresponding to frequencies smaller than the 
expansion rate at the time of production, $f_*=\HH_*/(2\pi)$, 
\begin{equation}\label{e:freqToday}
f_* = \frac{1}{2\pi\eta_*} \approx 10^{-8}\left(\frac{T_*}{\rm GeV}\right)\,\,{\rm Hz}~.
\end{equation}
For the EW scale this corresponds to $f_*^{\rm EW} \sim 10^{-6}$ Hz, while for the GUT 
scale the associated frequency is $f_*^{GUT} \sim 10^{8}$ Hz. For a given
energy scale $M\simeq T_*$ at the time of production, we are describing one 
frequency range or another, but always frequencies smaller than the one 
corresponding today to that energy scale, $f < f_*(M) \sim 10^{-8}{\rm Hz}(M/{\rm GeV})$. 
Clearly, only processes taking place in the radiation dominated Universe generate GWs with sufficiently high frequencies such that they can be observed by direct GW detection experiments. Indeed the frequency associated to the horizon at the matter-radiation equality is 
far too small, $f_*^{\rm eq} \sim 10^{-17}$ Hz, to be observed by
 direct GW detectors, like LIGO, LISA or BBO will be working.  Therefore
we consider only processes in the radiation dominated Universe and $\gamma = 1$ and 
$\nu = 2$ are assumed for the rest of the paper.

\section{The gravitational wave spectrum today}\label{s:GWtoday}

In this section we study two different cases, first the situation in which the source producing GWs lasts only a small fraction of the Hubble time at the moment of production and, second, 
the case in which the GW source acts for a much longer time, until the moment at which a
given mode enters the horizon. 

\subsection{Short lived source}

We first estimate the amplitude of the GW spectrum for large wavelengths, 
$k< \HH_*$, from a short lived source which lasts from $\eta_*$ to 
$\eta_{\rm fin}$, such that $( \eta_{\rm fin}-\eta_*)/\eta_{*} 
\equiv \epsilon \ll 1$ (as {\it e.g.} for the radial mode of $\phi$ in hybrid 
preheating~\cite{Bellido1,Bellido2}). Let us first note the following facts:\\ \\
1) From Eq.~(\ref{spectrumGW}) we see immediately that for small 
wavenumbers, $k\eta_{\rm fin}\ll 1$, the result scales like 
$$\frac{d\rho_{\rm GW}}{d\log k} \propto k^3 \ .$$
2) Since the source is short lived, $\eta_* \approx \eta_{\rm fin}$, and we deal with 
superhorizon modes, $k\eta_* \ll 1$, we may set $\cos(k\eta-k\eta') 
\approx 1$ and the time integral can be replaced simply by a factor 
$\ep\eta_*$.
\\ \\
3) To estimate the momentum integral, we use that Bessel functions at small
arguments, $x \equiv k\eta < 1$, can be approximated by  
$J_{\nu}(x) \approx (x/2)^{\nu}/\Gamma(1+\nu)$. To obtain the
dominant contribution at large wavelength ({\it i.e.} the least blue part) we
may also set $|\bk-\bq|\eta_* \simeq q\eta_*$.

\

Using all the above considerations, we are left with a simple integral for 
the evaluation of the spectra of the IR modes ($k\eta_*\ll1$) of GWs, 
at any time $\eta \gg \eta_*$ for which those modes have already crossed the 
horizon
\begin{eqnarray}
&& \left.\frac{d\rho_{\rm GW}(\eta)}{d\log k}\right|_{k\eta_*\ll1} \simeq 
	\frac{Gv^4}{4\pi^4} 36\pi^4A^2 \frac{k^3}{a^4(\eta)}\frac{2\pi}{N}\int_{-1}^1 d\cos\theta\sin^4\theta 
	\int_{0}^{1/\eta_*} dp\,\frac{p^6}{2^{2\nu}\Gamma^4(\nu+1)}\,
	\nn \\ &&\qquad  \times
	\left(\int_{\eta_*}^{\eta_{\rm fin}}d\tau\,a(\tau)\tau^3\right)^2 
	= \frac{3\cdot5 \pi^3}{7\cdot2^{11}}\frac{Gv^4}{N}\left({a_*\over a(\eta)}\right)^4\ep^2 H_*^2\,(k\eta_*)^3\,,
\end{eqnarray}
where we used $A = 5\pi/4$, $\nu = 2$ and we approximated $\int_{\eta_*}^{\eta_{\rm fin}}d\tau a(\tau)\tau^3 \approx a(\eta_*)\eta_*^3\delta\eta_* = \epsilon\,a(\eta_*)\eta_*^4$, since
we have set $\eta_{\rm fin}-\eta_* =\delta\eta_* \simeq \ep \eta_*$.

With this we can now evaluate the ratio of the GW energy density to the 
critical density today, for the IR modes $k\eta_*\ll1$,  as
\begin{eqnarray}
\Omega_{\rm GW}(f) &=& \frac{1}{\rho_c}
	 \frac{d\rho_{\rm GW}(\eta_0)}{d\log k} \approx 
	\frac{5\pi^4}{7\cdot 2^{8}}\left(\frac{v}{M_{\rm Pl}}\right)^4{\ep^2\over N}\,\Om_{\rm rad}(k\eta_*)^3 
	\nn \\ 
	&\sim& 10^{-5}\left(\frac{v}{M_{\rm Pl}}\right)^4{\ep^2\over N}\,(k\eta_*)^3  ~,
\label{e:OmGW}
\end{eqnarray}
where we used $H_*^2 = 8\pi G\rho_*/3$, we expressed the radiation density today as $\rho_{\rm rad} \approx \rho_*(a_*/a_0)^4$ and we introduced the the radiation density 
parameter today as $\Omega_{\rm rad} \approx 4.2\times10^{-5}$. We have also neglected the factors coming from the ratio of the 
effective relativistic degrees of freedom since they appear only with the
power $1/3$. 
\\

Note that this formula is general for the IR spectrum of GWs generated at any process 
in which the source, a $N$-component scalar field, has rapidly
acquired its true {\em vev} $v$ at $\eta_*$ and undergoes a short phase of
self-ordering which lasts for a fraction $\epsilon < 1$ of the Hubble time.
\\

Finally, note also that very generically we have $\eta_* \propto T_*^{-1} \propto 1/v$ so that
$\Om_{\rm GW} \propto v^4\eta_*^3k^3 \propto v\,k^3$ and not as $v^4$, as one could 
naively have concluded from Eq.~(\ref{e:OmGW}).

\subsubsection{The electroweak phase transition}

The comoving horizon size at the  electroweak (EW) phase transition 
is given by the EW energy scale $T_{*} \sim 100$ GeV,
$g_{\rm eff}(T_*) = 106.75$,
\begin{eqnarray*}
\eta_* \simeq 7.5 \times 10^4\ {\rm sec} \ .
\end{eqnarray*}
Inserting this above with $f=k/(2\pi)$, we find
\begin{eqnarray}
\Omega_{\rm GW}(f) \approx 4.2\times 10^5\,\frac{5\pi^4(2\pi)^3}{7\cdot2^{8}}\,\Om_{\rm rad}
\left(\frac{v}{M_{\rm Pl}}\right)^4 {\ep^2\over N}
\left(\frac{f}{\rm mHz}\right)^3 \sim 10^{-65}
\,{\ep^2\over N}\left(\frac{f}{\rm mHz}\right)^3 \ .
\end{eqnarray}
For the last expression we have used $v\simeq T_*$. This result is of course unmeasurably small.

\subsubsection{A GUT scale phase transition}

To have any chance to measure this spectrum, we need a \textit{vev} which is not too
many orders of magnitude below that Planck scale, since the GW energy density is suppressed by a fourth 
power of the ratio of the $vev$ to $M_{\rm Pl}$. The best change might be 
a GUT scale with a \textit{vev} of the order of $v\simeq 10^{16}$GeV. But then
of course $\eta_*$ will be very small and the dominant contribution will come 
from very high frequencies, lower frequencies being suppressed by the factor
$(k\eta_*)^3$. For $T_* =10^{16}$GeV we have
$$ \eta_* \simeq 5\times 10^{-10}\ {\rm sec} \ , $$
leading to 
\begin{eqnarray}
\Omega_{\rm GW}(f) \approx 0.125\,\frac{5\pi^4(2\pi)^3}{7\cdot2^{8}}\,\Om_{\rm rad}
\left(\frac{v}{M_{\rm Pl}}\right)^4 {\ep^2\over N}
\left(\frac{f}{\rm GHz}\right)^3 \sim 10^{-16}\,
{\ep^2\over N}\left(\frac{f}{\rm GHz}\right)^3 \ .
\end{eqnarray}
Apart from the fact that this result suffers severe additional suppression at 
measurable frequencies which are signi\-ficantly below 1GHz $=10^9$Hz,
the sensitivity of $10^{-12}\Om_{\rm rad} \simeq 10^{-16}$ cannot be reached 
with any presently proposed experiment at those frequencies.

\

Therefore, we can only conclude that the superhorizon GW spectrum generated from a short 
lived self ordering scalar field is much below presently proposed experimental sensitivities.

\subsection{A long lived source}

As we have seen in the previous subsection, short lived Goldstone modes 
cannot lead to a significant GW background. But since
Goldstone modes are typically non-interacting and long lived, it is more
natural to consider them for a time which is much longer than the
horizon scale $\eta_*$.
To compute the GW energy density produced by such
a self ordering scalar field, we consider Eq.~(\ref{spectrumGW}) 
and set $\eta_{\rm fin} = \eta_k \equiv 1/k$, since the solution (\ref{e:beta2}) decays inside 
the horizon, when $k\eta>1$.
We then have to compute the following integral 
\bea  \label{e:long1}
&&  \frac{d\rho_{\rm GW}(k,\eta_k)}{d\log k}= \frac{G\,v^4}{4\pi^4}
    \frac{k^3}{a^4(\eta_k)}\frac{36\pi^4A^2}{N}
    \int_{\eta_*}^{1/k} d\tau 
     \int_{\eta_*}^{1/k} d\xi \
     a(\tau) a(\xi) \cos(k\xi-k\tau) \times
     \nn\\ && \qquad
      \int_{\begin{tiny}\begin{array}{c} p\eta_*<1\\ |\bp-\bk|\eta_*<1 \end{array}
           \end{tiny}}  d^3p~p^4 \sin^4\theta ~
     \tau^3 \xi^3 \
     \frac{J_{\nu}(p\tau)}{(p\tau)^{\nu}} 
     \frac{J_{\nu}(p\xi)}{(p\xi)^{\nu}}
     \frac{J_{\nu}(|\bk-\bp|\tau)}{(|\bk-\bp|\tau)^{\nu}}
     \frac{J_{\nu}(|\bk-\bp|\xi)}{(|\bk-\bp|\xi)^{\nu}}
     ~,
\eea
Note that the range of integration of the variable $p$  in the above expression is set to be $\{p\eta_*<1,\ |\bp-\bk|\eta_*<1\}$ since the initial two point correlator of the scalar field turns out to be different from zero only in this range of momenta [{\it c.f.} Eq.~(\ref{e:coreta*})].

In order to obtain an analytical result for the above
integral, we perform the following approximations:
\begin{itemize}
\item
  We are interested in scales $k$ that are superhorizon for all the time of
  GW production, namely $k\tau < 1$ and $k\xi < 1$ for times 
  $\tau, \,\xi$ between $\eta_*$ and $\eta_{\rm fin}=1/k$, therefore we approximate $\cos(k\xi-k\tau) \simeq 1$  .
\item
  We neglect the angular dependence of $|\bp-\bk|$ so that the angular
  integral reduces to $2\pi\int\sin^4\theta d\cos\theta = 32\pi/15$.
\item
  In the range of integration where $p\tau\gg 1$ we substitute
  $|\bk-\bp|\tau \simeq p\tau$, while when $p\tau\ll 1$ we 
  approximate  $|\bk-\bp|\tau \ll 1$.
\item
The range of momenta for which we can expand the Bessel functions in terms of small arguments is 
$p < {\rm min}(1/\tau,1/\xi)$, while in the range ${\rm min}(1/\tau,1/\xi)<p<{\rm max}(1/\tau,1/\xi)$ we 
should distinguish between large and small argument expansions of the Bessel functions. 
Finally, in the range  ${\rm max}(1/\tau,1/\xi)<p<1/\eta_*$ one can consider the 
large argument limit for all the four Bessel functions of the above integral.
\end{itemize}

Taking into account all the above considerations, we find that the complete integral becomes
$$
\int_{\eta_*}^{1/k}\!d\tau\,\int_{\eta_*}^{1/k}\! d\xi\,\int_0^\infty\!dp\, f(p,\tau,\xi) \ = \
2\int_{\eta_*}^{1/k}\! d\tau\,\int_{\eta_*}^\tau\!d\xi\,\left(\int_0^{1/\tau}\!dp\, f+ 
\int_{1/\tau}^{1/\xi}\!dp \,f+ \int_{1/\xi}^{1/\eta_*}\! dp \,f\right)\,,
$$
which allows us to separate the integral in $p$ using the asymptotic behaviour of the Bessel functions, 
\bean
&&  J_\nu(x) \simeq \frac{x^\nu}{2^\nu \Gamma(\nu+1)}  \hspace{3.7cm} {\rm for }~ x\ll 1~,
\\
&&  J_\nu(x) \simeq \sqrt{\frac{2}{x\pi}}\cos\lp(x -\frac{(2\nu+1)\pi}{4}\rp)  \hspace{1cm} {\rm for }~ x\gg 1~.
\eean
We can distinguish three different intervals:
\begin{itemize}
\item
The IR contribution, $I_1(k)$,  for $0<p<1/\tau$, with $|\bk-\bp|\tau < 1$ and $|\bk-\bp|\xi < 1$\,.
\item
The mixed (UV+IR) contribution, $I_2(k)$, for $1/\tau<p<1/\xi$, with $|\bk-\bp|\tau \simeq p\tau > 1$ but
 $|\bk-\bp|\xi \simeq p\xi < 1$\,. 
 \item
The UV contribution, $I_3(k)$, for $1/\xi<p<1/\eta_*$, with $|\bk-\bp|\tau \simeq p\tau > 1$ and 
   $|\bk-\bp|\xi \simeq p\xi > 1$\,.
\end{itemize}   

Therefore we can finally write
  \be  \label{e:long2}
  \frac{ d\rho_{\rm GW}(k,\eta_k)}{d\log k} =
  	\mathcal{D}(k) \lp[I_1(k)+I_2(k)+I_3(k)\rp]~,
  \ee	
  where the pre-factor $\mathcal{D}(k)$ contains the coefficients in front of the integral in Eq.~(\ref{e:long1}), the factor coming from the angular 
  integration ($32\pi/15$) and  the factor 2 that comes from the symmetry of the double time integration, namely
  \be 
  \mathcal{D}(k) \equiv  \frac{G\,v^4}{4\pi^4}    \frac{k^3}{a^4(\eta_k)}\frac{36\pi^4A^2}{N}
		\times\frac{32\pi}{15}\times 2 = \frac{G\,v^4}{N}\frac{k^3}{a^4(\eta_k)}15\cdot 4\pi^3~. \ \ \
  \ee
The three integrals of Eq.~(\ref{e:long2}) are given by
\bea
I_1(k) &\equiv& \int_{\eta_*}^{1/k}\!d\tau \int_{\eta_*}^{\tau}\!d\xi\ a(\tau)\ a(\xi)\ \tau^3\ \xi^3 
	\int_0^{1/\tau}dp  ~p^6 \
     \frac{J_{\nu}(p\tau)}{(p\tau)^{\nu}} 
     \frac{J_{\nu}(p\xi)}{(p\xi)^{\nu}}
     \frac{J_{\nu}(|\bk-\bp|\tau)}{(|\bk-\bp|\tau)^{\nu}}
     \frac{J_{\nu}(|\bk-\bp|\xi)}{(|\bk-\bp|\xi)^{\nu}}    
     \nn \\ 
     &\simeq &  \frac{H_0^2\Omega_{\rm rad}}{4096}  \int_{\eta_*}^{1/k}\!d\tau 
     \int_{\eta_*}^{\tau}\!d\xi ~ \tau^4\ \xi^4
     \int_0^{1/\tau} dp\ p^6 \nn \\
     &=& \frac{H_0^2\Omega_{\rm rad}}{4096\ k^3} \frac{1}{35} \left[\frac{1}{3}-
     \frac{5}{6}(k\eta_{*})^3+ \frac{1}{2}(k\eta_{*})^5 \right]  \,,
\eea
\bea
I_2(k) &\equiv& \int_{\eta_*}^{1/k}\!d\tau \int_{\eta_*}^{\tau}\!d\xi\ a(\tau)\ a(\xi)\ \tau^3\ \xi^3 
	\int_{1/\tau}^{1/\xi}dp  ~p^6 \
     \frac{J_{\nu}(p\tau)}{(p\tau)^{\nu}} 
     \frac{J_{\nu}(p\xi)}{(p\xi)^{\nu}}
     \frac{J_{\nu}(|\bk-\bp|\tau)}{(|\bk-\bp|\tau)^{\nu}}
     \frac{J_{\nu}(|\bk-\bp|\xi)}{(|\bk-\bp|\xi)^{\nu}}    
     \nn \\ 
     &\simeq &  \frac{H_0^2\Omega_{\rm rad}}{32\pi} \int_{\eta_*}^{1/k}\! d\tau \int_{\eta_*}^{\tau}\!d\xi\ \tau^4\ \xi^4
     \int_{1/\tau}^{1/\xi}\frac{dp\ p^6}{(p\tau)^{5}}\ \cos^2\lp(p\tau - {5\pi\over4}\rp)
     \nn \\
     &=& \frac{H_0^2\Omega_{\rm rad}}{128\pi \ k^3} \left[\frac{2}{45}+\frac{1}{18}(k\eta_*)^3- 
   \frac{1}{10}(k\eta_*)^5 +\frac{(k\eta_*)^3}{3} \log(k\eta_*) \right]  \,,
\eea
and
\bea   \label{int3}
I_3(k) &\equiv& \int_{\eta_*}^{1/k}\!d\tau \int_{\eta_*}^{\tau}\!d\xi\ a(\tau)\ a(\xi)\ \tau^3\ \xi^3 
	\int_{1/\xi}^{1/\eta_*}dp  ~p^6 \
     \frac{J_{\nu}(p\tau)}{(p\tau)^{\nu}} 
     \frac{J_{\nu}(p\xi)}{(p\xi)^{\nu}}
     \frac{J_{\nu}(|\bk-\bp|\tau)}{(|\bk-\bp|\tau)^{\nu}}
     \frac{J_{\nu}(|\bk-\bp|\xi)}{(|\bk-\bp|\xi)^{\nu}}    
     \nn \\ 
     &\simeq &  \frac{4\,H_0^2\Omega_{\rm rad}}{\pi^2}  \int_{\eta_*}^{1/k}\!d\tau \int_{\eta_*}^{\tau}\!d\xi\ \tau^4\ \xi^4
     \int_{1/\xi}^{1/\eta_*}\frac{dp\ p^6}{(p\tau)^{5}(p\xi)^{5}}  \cos^2\lp(p\tau - \frac{5\pi}{4}\rp)
      \cos^2\lp(p\xi - \frac{5\pi}{4}\rp)
     \nn \\
     &=& \frac{H_0^2\Omega_{\rm rad}}{3\pi^2\ k^3}  \left[\frac{1}{9}-\frac{1}{9}(k\eta_*)^3- 
     (k\eta_*)^3\left( \frac{1}{2} \log^2(k\eta_*)-\frac{1}{3} \log(k\eta_*)\rp) \right]  ~.
\eea
More precisely, in the above computation we substituted each $\cos^2 x$ by its mean value 
$\left\langle\cos^2 x\right\rangle = 1/2$ averaged over
a few oscillations, and we introduced the usual expression for the scale factor in a 
radiation dominated background, $a(\eta) \simeq H_0 \sqrt{\Omega_{\rm rad}} \eta$, 
which is consistent with $a_0=1$ today.

All three terms have a scale-invariant spectrum.
Actually, the "UV" contribution given in Eq.~(\ref{int3}) is the largest. Summing all the three contribution and 
considering the dominant part in the limit $k\eta_* \ll1$ [hence also 
$(k\eta_*)^3 \log(k\eta_*) \ll 1$], we obtain the following scale-invariant spectrum
\bea
  \frac{ d\rho_{\rm GW}(k,\eta_k)}{d\log k} &\simeq&  5\cdot2^5 \pi^4
 \frac{\Om_{\rm rad}\, \rho_c}{Na^4(\eta_k)} \lp(\frac{v}{M_{\rm Pl}}\rp)^4 
    \left( \frac{1}{2^{12} \cdot 105} + \frac{1}{2^6\pi \cdot 45} + \frac{1}{27\pi^2}\right)
    \nn\\
     &\simeq& 60\times\frac{\Omega_{\rm rad}\,\rho_c}{Na^4(\eta_k)} \lp(\frac{v}{M_{\rm Pl}}\rp)^4  ~,
\eea
where  we have used the Friedmann equation $H^2_0=8\pi G \rho_c/3$.
Redshifting the above expression until today, we obtain for the GW 
energy density parameter,
\be
	\label{OmegaGWlong}
  \Omega_{\rm GW}(k,\eta_0) \equiv
   \frac{ d\rho_{\rm GW}(k,\eta_0)}{ \rho_c d\log k} =
  \frac{d\rho_{\rm GW}(k,\eta_k)}{\rho_c d\log k} 
    a^4(\eta_k)  \simeq \frac{60}{N}~\Omega_{\rm rad} \lp(\frac{v}{M_{\rm Pl}}\rp)^4   \,.
\ee
This corresponds to a scale-invariant GW spectrum produced by
a self-ordering scalar field in the large $N$-limit. This result is valid for
all wave numbers $k$ which enter the horizon when the Goldstone modes of our
$N$-component field are still massless and the field has not yet decayed.
Scales which enter the horizon after this time $\eta_{\rm fin}$, $i.e.$ scales 
with $k\eta_{\rm fin} <1$, are suppressed by a factor  $(k\eta_{\rm fin})^3$,
as for them the result for a short lived source with $\eta_*$ replaced by 
$\eta_{\rm fin}$ applies.

\FIGURE[ht]{ 
	\epsfig{width=15cm,height=9cm, file=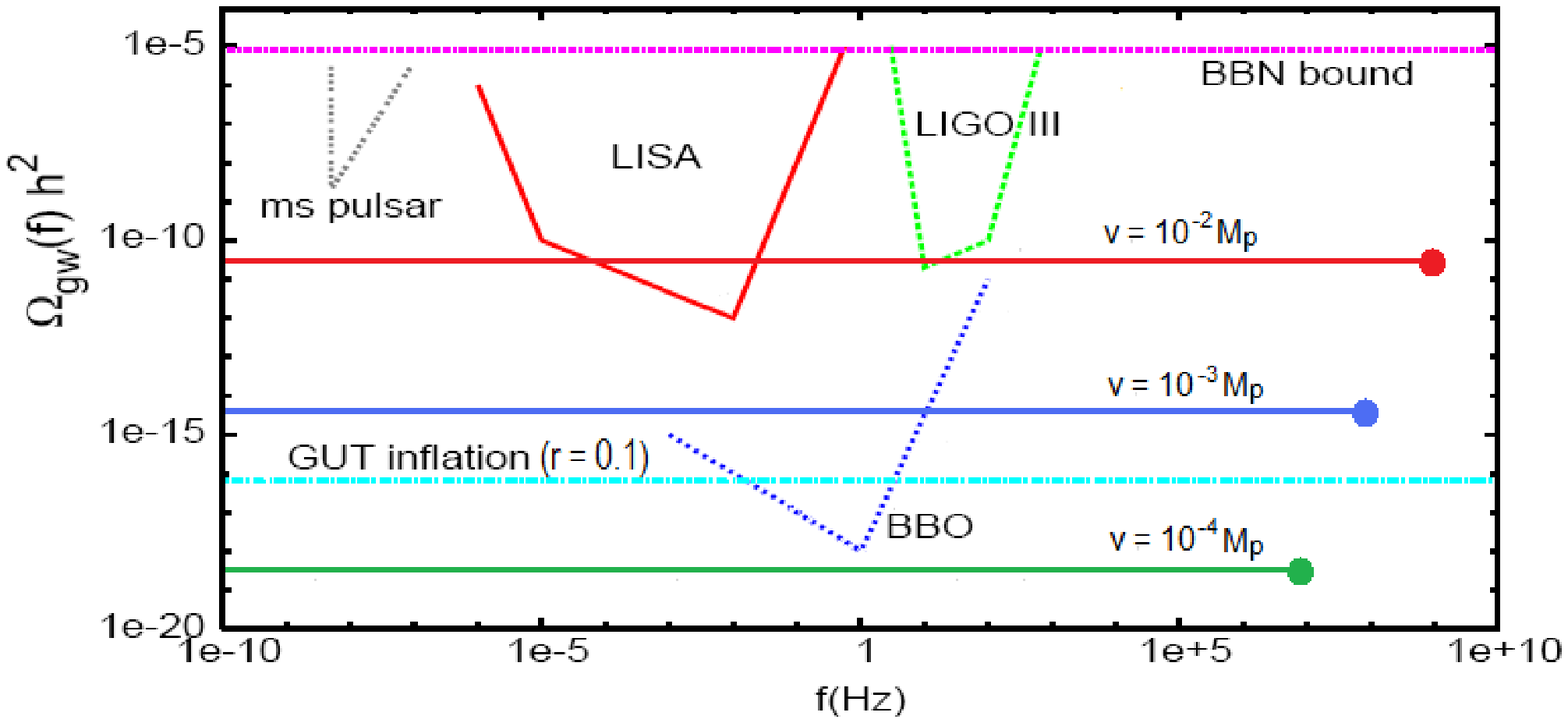}
\caption{The sensitivity of present and future GW experiments are 
compared with our results for a long lasting source and inflation. We show, 
the amplitude of the scale-invariant GW background expected from a GUT scale 
inflation (blue, dashed) and from a self-ordering long lived source as 
studied in this paper, for a symmetry breaking field with $N = 4$ real 
components and a $vev$ $v = 10^{-2} M_{\rm Pl}$ (top, red line), 
$v = 10^{-3} M_{\rm Pl}$ (middle, blue line, overlying with inflation) and 
$v = 10^{-4} M_{\rm Pl}$ (bottom, green line). The big dot at the 
right end of the horizontal lines represents the frequency~(\ref{e:freqToday}) 
associated to the horizon at the initial time of production. }
	\label{fig1}
	}    

\FIGURE[ht]{ 
	\epsfig{width=13cm, file=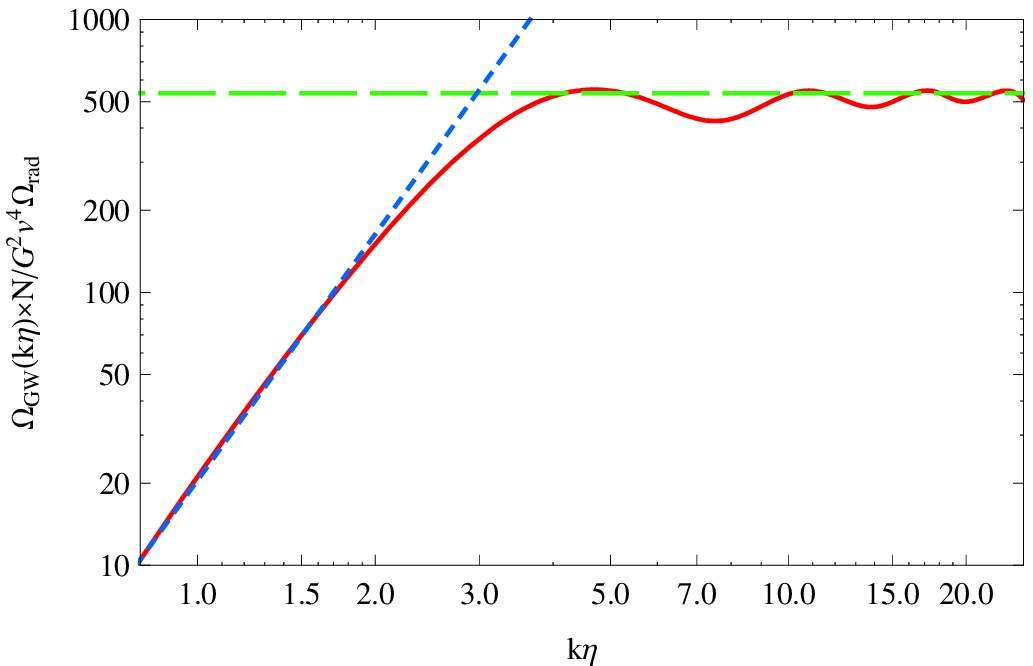}
\caption{The density parameter in gravitational waves as a function of $k\eta$.
For scales outside the horizon, $k\eta<\pi$, we observe the $(k\eta)^3$ dependence
(short dashed line), while for scales that have entered the horizon, $k\eta>\pi$, the 
GW energy density saturates, at a normalized value of 511 (long dashed line). 
This result implies a significant scale-invariant GW spectrum today.}
	\label{fig2}
	}

\subsection{Numerical integration}

In order to obtain more accurate results, and to check the validity of our 
analytical approximations, we have also performed a numerical 
evaluation of the integrals in Eq.~(\ref{e:masterEq}). If we set the final 
time of integration to be the horizon 
crossing, $\eta_{\rm fin}=1/k$, as we did in the analytical evaluation 
for the long lasting source~(\ref{e:long1}), we obtain the following
result for the final GW density parameter today
\be
   \Omega_{\rm GW}(k,\eta_0) \simeq \frac{22}{N} \Omega_{\rm rad} 
 \lp(\frac{v}{M_{\rm Pl}}\rp)^4~ \, , \qquad \eta_{\rm fin}=1/k \, .
\ee
This suggests that the analytical approximation somewhat overestimates 
the result.  However, we can continue the integration to later times when 
the wavelength has already entered the horizon.

The integral in Eq.~(\ref{e:long1}) allows us to compute the GW energy density in the
limit $k\eta_*\ll1$, using the change of variables $u=\cos\theta,~q=p/k,~x=k\tau$,
\be  \label{OMGW}
\Omega_{\rm GW}(k,\eta) = {G^2v^4\Omega_{\rm rad}\over Na^4(\eta)}~75\pi^4
\hspace{-1mm}\int_0^\infty  \hspace{-2mm}dq~q^2 F(q)  \left\{\left[\int_0^{k\eta} \hspace{-2mm}
dx\,\cos x\,J_2^2(q x)\right]^2\hspace{-2mm}+
  \left[\int_0^{k\eta} \hspace{-2mm}dx\,\sin x\,J_2^2(q x)\right]^2\right\}
\ee
where the kernel $F(q)$ comes from the integration over angles,
\be \nn
F(q) = \hspace{-1mm}\int_{-1}^1  {du\,(1-u^2)^2\over(q^2+1-2qu)^2} 
  ={1\over24 q^5} \left[16q + 12q(q^2-1)^2+ 
   3 (q^2-1)^2 (q^2+1) \log{(q-1)^2\over(q+1)^2}\right]
\ee
and we have made the approximation, $J_2(x\sqrt{q^2+1-2qu})
\rightarrow J_2(q x)$, inside the time integration. We have checked that for
large times the result is correct within 0.1\%.

Numerically evaluating  (\ref{OMGW}), we find that the GW energy density 
continues to grow until horizon crossing, $k\eta \simeq \pi$, and saturates 
thereafter, see Fig.~\ref{fig2}. 
This agrees with the result of Ref.~\cite{JonesSmith:2007ne}, who find a peak 
in the power spectrum $\PP(k,\eta)$ at approximately this value, and also 
explains the $1/a(\eta)^2$ dependence of the Power spectrum, $\PP\propto
\Omega_{\rm GW}/a^2$, for scales that have already entered the horizon.

For $k\eta \gg 4$ the gravitational wave energy density saturates at a value
\be
\label{e:omegaFin}
   \Omega_{\rm GW}(k,\eta_0) \simeq \frac{511}{N} \Omega_{\rm rad} 
   \lp(\frac{v}{M_{\rm Pl}}\rp)^4 \,,
\ee 
where we used again the usual normalization of the scale factor in a radiation 
dominated background.
These results suggest that the GW spectrum produced by this mechanism still 
grows inside the horizon and reaches its final value somewhat after horizon 
crossing. This is consistent with the fact that the power of the scalar field 
that sources these GWs is not absent inside the horizon, but it is indeed 
given by the Bessel functions in Eq.~(\ref{e:Pbeta}), which decay rather
slowly as functions of $k\eta$. 

 In the following analysis we will consider the numbers arising from the 
numerical integration, as given in Eq.~(\ref{e:omegaFin}).

\subsection{Observational constraints}

Our result for the amplitude of the GW spectrum~(\ref{e:omegaFin}) is inside 
the range of detectability of
the BBO~\cite{bbo} experiment ($\Omega_{\rm GW}(k)\gtrsim 10^{-17}$)  and 
is marginally detectable by LISA~\cite{lisa} or advanced LIGO~\cite{ligo} 
($\Omega_{\rm GW}(k)\gtrsim 10^{-10}$). 
Indeed, with $\Omega_{\rm rad} \simeq 4.2 \times 10^{-5}$, we find that BBO 
would detect this signal if the symmetry breaking scale $v$ satisfies
\[
   \lp(\frac{v}{M_{\rm Pl}}\rp)^4 \gtrsim 4.7 \cdot10^{-16} N 
      \qquad \Rightarrow \qquad 
   \frac{v}{M_{\rm Pl}} \gtrsim 1.5 \cdot10^{-4} N^{1/4} ~.
\]
Concerning the sensitivity of LIGO or LISA, the signal is detectable if
\[
   \lp(\frac{v}{M_{\rm Pl}}\rp)^4 \gtrsim 4.7 \cdot10^{-9} N 
      \qquad \Rightarrow \qquad 
   \frac{v}{M_{\rm Pl}} \gtrsim 0.008 \,N^{1/4} ~.
\]
In other words, for scales higher or around the GUT scale, $v \gtrsim 10^{16}$GeV, the 
very long wavelength tail which we have studied here could be observed.

In order to relate the above scale-invariant GW energy density to the GW  
spectrum from inflation, we compute the relative tensor-to-scalar ratio $r$. 
Following Ref.~\cite{Chongchitnan:2006pe}, one has the following expression for 
the GW density parameter from inflation 
\be
  \Omega_{\rm GW}(k,\eta_0) = 4.36 \times 10^{-15} \,r \lp(\frac{k}{k_0}\rp)^{n_{\rm T}} \,, 
  \qquad  r\equiv \frac{\PP_{\rm T}(k_0)}{\PP_{\rm S}(k_0)}~,
\ee
where $k_0=0.002\,h\,$Mpc$^{-1}$, $\PP_{\rm T}(k)=
   r\PP_{\rm S}(k_0)(k/k_0)^{n_{\rm T}}$  
and we used the WMAP result, $\PP_{\rm S}(k_0) = 2.21 \times 10^{-9}$. 
This concerns only the wavelengths which enter the horizon in the radiation 
dominated era, before equality. Comparing the above expression for 
$n_T\simeq 0$ with our Eq.~(\ref{e:omegaFin}), we obtain in our case
\be
  r\simeq \frac{3}{N}  \lp(\frac{v}{10^{16}{\rm GeV}}\rp)^4  .
\ee
Another usefull comparison with inflation is the relative strength of the GW 
energy densities produced by the above two different mechanisms.
Considering always wavelengths which enter the horizon in the radiation 
dominated epoch, we have \cite{Maggiore:1999vm}
\be
\Om_{GW}^{(\rm inf)} = 10^{-13} \lp(\frac{H_*}{10^{-4} M_{\rm Pl}}\rp)^2 = 
  8.4 \times 10^{-5} \lp(\frac{M}{M_{\rm Pl}}\rp)^4\,,
\ee
where $M$ denotes the energy scale of inflation, $H_*^2 \equiv 8\pi G M^4/3$.
The ratio between the GW energy density produced by our mechanism and the 
one from inflation is then
\be
\label{e:R}
 \mathcal{R}\equiv 
\frac{ \Omega_{\rm GW}(k,\eta_0)}{\Omega_{\rm GW}^{(\rm inf)}(k,\eta_0)}
  	\simeq \frac{256}{N}\lp( v\over M\rp)^4 \,. 
\ee

Comparing these results with those of Ref.~\cite{JonesSmith:2007ne}, where the 
authors mainly concentrate on the spectrum of GWs produced in a matter 
dominated universe, we reproduce perfectly the amplitude of their spectrum 
$\PP(k,\eta)$ defined as in Eq.~(\ref{def:P}), but their final 
relative strength $\RR$ is nearly 2 orders of magnitude larger than what we 
find in Eq.~(\ref{e:R}). We believe this is due to the factor $1/(2\pi^3)$ missing in 
their expression for $\Omega_{\rm GW}(k, \eta_0)$ which has to be introduced 
for consistency with the definition of the power spectrum $\PP(k)$.

\section{Conclusions}\label{s:con}

In this paper we have estimated the contributions to the gravitational 
wave background from a symmetry breaking phase transition on large
scales, $k\eta_*<1$. We have concentrated on the analysis of the Goldstone 
modes and we obtained the following main conclusions.

If the modes are {\em short lived} with duration $\ep\eta_*$, $\ep<1$ their 
contribution is blue and suppressed by a factor $\ep^2(k\eta_*)^3$. This 
result is actually generic, independent of the
nature of the short lived source. Indeed, one typically obtains
\be\label{e:OmGWsl}
  \Omega_{\rm GW}(k) \simeq (k\eta_*)^3
  \Omega_{\rm rad}\Om_X^2\ep^2 \ ,
\ee
where $\Om_X$ is the density parameter of the source of anisotropic stresses 
at the moment of creation.  For the Goldstone modes the factor $\Om_X^2$ is
replaced by $(v/M_{\rm Pl})^4$. This strong suppression factor renders
GWs from short-lived Goldstone modes entirely unobservable.

The situation is different for {\em long lived} Goldstone modes.
There the suppression factor $(k\eta_*)^3$
is absent. Therefore, if the Goldstone modes remain massless until a time 
$\eta_{\rm fin}$, for modes with $k\eta_{\rm fin}\gsim 1$ the spectrum is 
scale invariant and the amplitude is given by 
\be\label{e:OmGWll}
  \Omega_{\rm GW}(k) \simeq \frac{511}{N} 
  \Omega_{\rm rad}\lp(\frac{v}{M_{\rm Pl}}\rp)^4 \ ,
\ee
which is marginally detectable with the experimental sensitivity of advanced 
LIGO or LISA and is well within the range of BBO  for a GUT scale phase 
transition. The results  for the long-lived source are  summarized in Fig.~\ref{fig1}.

If the Goldstone modes are still present at decoupling, $\eta_{\rm fin}\gsim 
\eta_{\rm dec}$, these GWs will also leave a signature in the cosmic 
microwave background where they lead to a scale-invariant contribution
very similar to the one of global textures, i.e. a $N=4$ global $\OO(N)$
model~\cite{DKM}.

Note that this new GW background from self-ordering fields after inflation 
(e.g. from hybrid preheating) has a power spectrum very similar to that coming 
from inflation, and therefore it may become important  to disentangle both if they 
are present simultaneously, that is if the scale of inflation and that of symmetry
breaking are related by parameters of order one, like in hybrid inflation.

\section*{Acknowledgments} 
DGF would like to express his gratitude to the Cosmology 
Group of Geneva University, for hospitality received during the fall 2008, 
when this project was initiated. EF is grateful to the Universidad 
Aut\'onoma de Madrid for hospitality during the spring of 2009. JGB thanks 
the Theory Group at CERN for hospitality during the summer of 2009. 
DGF also acknowledges support from 
a FPU-Fellowship from the Spanish M.E.C., with Ref.~AP-2005-1092. 
This work is supported by the Swiss National Science Foundation.
We also acknowledge financial support from the Madrid Regional Government 
(CAM) under the program HEPHACOS P-ESP-00346, and the Spanish Science
Research Ministry (MEC) under contract FPA2006-05807. 
DGF and JGB participate in the Consolider-Ingenio 2010 PAU (CSD2007-00060), 
as well as in the European Union Marie Curie Network 
``UniverseNet" under contract MRTN-CT-2006-035863.

\end{document}